\documentclass[iop]{emulateapj}
\usepackage{epsfig}
\usepackage{amsmath}
\usepackage{natbib}
\usepackage{apjfonts}

\begin{document}

\title{Numerical Modeling of the 2009 Impact Event on Jupiter}
\author{Jarrad W.\ T.\ Pond\altaffilmark{1}} 
\author{Csaba Palotai\altaffilmark{1}}
\author{Travis Gabriel\altaffilmark{1}}
\author{Donald G.\ Korycansky \altaffilmark{2}}
\author{Joseph Harrington\altaffilmark{1}}
\author{No\'{e}mi Rebeli\altaffilmark{1}}
\email{jarradpond@gmail.com}

\affil{$^{1}$ Planetary Sciences Group, Department of Physics, University of Central Florida, Orlando, FL 32816-2385, USA}
\affil{$^{2}$  Department of Earth and Planetary Science, University of California, Santa Cruz, CA 95064, USA}
\date{\today}


\begin{abstract}
We have investigated the 2009 July impact event on Jupiter using the ZEUS-MP 2 three-dimensional hydrodynamics code. We studied the impact itself and the following plume development. Eight impactors were considered: 0.5 km and 1 km porous ($\rho = 1.760$ g cm$^{-3}$) and non-porous ($\rho = 2.700$ g cm$^{-3}$) basalt impactors, and 0.5 km and 1 km porous ($\rho = 0.600$ g cm$^{-3}$) and non-porous ($\rho = 0.917$ g cm$^{-3}$) ice impactors. The simulations consisted of these bolides colliding with Jupiter at an incident angle of $\theta = 69^{\circ}$ from the vertical and with an impact velocity of $v$ = 61.4 km s$^{-1}$. Our simulations show the development of relatively larger, faster plumes created after impacts involving 1 km diameter bodies. Comparing simulations of the 2009 event with simulations of the Shoemaker-Levy 9 events reveals a  difference in plume development, with the higher incident angle of the 2009 impact leading to a shallower terminal depth and a smaller and slower plume. We also studied the amount of dynamical chaos present in the simulations conducted at the 2009 incident angle. Compared to the chaos of the SL9 simulations, where $\theta \approx 45^{\circ}$, we find no significant difference in chaos at the higher 2009 incident angle.
\end{abstract}
 
\maketitle

\section{Introduction}
\label{sec:intro}
Between 1994 July 16 and 22, fragments of the comet D/Shoemaker-Levy 9 (hereafter SL9) penetrated the Jovian atmosphere. This predicted impact gained great attention, with many Earth- and space-based telescopes aimed at this spectacle (see \citealp{HarringtonEtal2004JupBook} for a review of the event). Despite impact occurrences of this nature being characterized as highly unlikely, another object collided with Jupiter sometime between UT 9-11 on 2009 July 19, drastically increasing the expected collision rate of 0.5-1.0 km bodies with Jupiter \citep{2010ApJ...715L.155S}. Unfortunately, the 2009 impact itself was not directly observed; however, it was analyzed through observations of the impact's aftermath and was compared to the SL9 impacts \citep{2010Icar..210..722D, 2010A&A...524A..46F, 2010ApJ...715L.150H, 2011Icar..211..587O, 2010ApJ...715L.155S}. Without a direct observation of the event, we use numerical simulations to seek a better understanding of the possible impact circumstances that could have produced this large atmospheric disturbance.

We use the ZEUS-MP 2 hydrodynamics code to simulate the collisions of several different types of impactors, sampling the impact parameter space constrained by observations. From simulation results, we garner information about possible penetration depths, plume development dynamics, and impact energies of potential 2009 impact scenarios. We also compare our results to numerical simulations of the SL9 impact event conducted by \cite{2006ApJ...646..642K} and \cite{2011ApJ...731....3P}. We compare energy deposition, penetration depth, and plume development between the SL9 and 2009 simulations.

This paper is organized as follows: Section \ref{sec:model} gives a description of our numerical model. The results of the simulations are given in Section \ref{sec:results}. Section \ref{sec:disc} discusses the results, compares them to observations, and also compares them to those of SL9 impact simulations. Lastly, the conclusions are contained in Section \ref{sec:con}.

\section{Impact Model}
\label{sec:model}
As with \cite{2006ApJ...646..642K} and \cite{2011ApJ...731....3P}, we used ZEUS-MP 2---a 3D, parallel hydrodynamics code---for impact simulations. See \cite{2006ApJS..165..188H} for ZEUS-MP 2 code details, and see \citet{2002Icar..157....1K} and  \citet{2003Icar..161..244K} for the modifications made to the ZEUS-MP code to include multiple materials. By extending both the simulation time and the spatial extent of the computational grid in which the simulations are run, our ZEUS-MP-based code can model both the impact phase and the entry-response/blowout phase of the impact (here, we follow the terminology of \citealp{HarringtonEtal2004JupBook}). Hence, the impactor's entry into the atmosphere, impactor break up, and plume formation and development are simulated in each run. 

In the present simulations, we use the same coordinate system and a similar computational grid as \cite{2011ApJ...731....3P}. A Cartesian coordinate system is used in the simulations: $x_{1}$, the ``along-track coordinate,'' is aligned with the impactor's initial trajectory; $x_{3}$, the ``cross-track coordinate," is perpendicular to the impactor's initial trajectory, and $x_{2}$, the horizontal axis, is perpendicular to both $x_{1}$ and $x_{3}$. The local Cartesian coordinates for the Jovian reference frame are given by $x$, $y$, and $z$ and are related to $x_{1}$, $x_{2}$, and $x_{3}$ by the following equations, given by \cite{2006ApJ...646..642K}:

\begin{eqnarray}
x &=& x_{2}, \nonumber \\
y &=& -x_{1}\sin{\theta} + x_{3}\cos{\theta}, \\
z &=& x_{1}\cos{\theta} + x_{3}\sin{\theta}, \nonumber
\end{eqnarray}
where $\theta$ is the angle of incidence, i.e., the angle between $x_{1}$ and the local vertical, $z$. The origins of both coordinate systems coincide with the location of the 1-bar pressure level in the simulated Jovian atmosphere. Within and in close proximity to the impactor, the resolution is constant at 16 grid cells across the radius of the impactor (R16). The grid spacing then increases geometrically in each direction away from the R16 area. The resolution is again held constant (4 km per grid cell) at the tail end of the grid, the area in which the impact plume develops and evolves.

As described by \cite{2006ApJ...646..642K}, we use the Tillotson equation of state (EOS). The Tillotson EOS was derived for cases requiring high-velocity impact calculations, can describe the transition of shocked material into the vapor phase, but cannot represent a two-phase region, i.e., where a liquid and gas co-exist \citep{1989icgp.book.....M}. The Tillotson EOS parameters used for the basalt and ice impactors are the same as those listed in Table 1 of  \cite{2006ApJ...646..642K}. These EOS parameters characterize the behavior of the different impactor materials.

For possible 2009 bolides, \cite{2010ApJ...715L.155S} suggest ice impactor diameters of $\sim$ 0.5 km up to $\sim$ 1 km based on comparisons to SL9 models and ablation rate considerations at higher angles of incidence. \cite{2010ApJ...715L.150H} suggest ice impactor diameters of 500-700 m based on similarities of the 2009 impact site to the E and R impact sites of SL9. There is observational evidence suggesting the possibility that the 2009 impactor was asteroidal in origin, rather than cometary \citep{2010A&A...524A..46F, 2010ApJ...715L.150H, 2011Icar..211..587O}. From thermal heating and mass transport estimates, \cite{2011Icar..211..587O} suggest diameters of 200-500 meters for basalt impactors of density 2.5 g cm$^{-3}$ (an impactor mass range of $\sim$ 1.05 $\times$ $10^{13}$ grams to $\sim$1.64 $\times$ $10^{14}$ grams). The following eight impact cases were run to $\sim$ 30 seconds after impact: 0.5 km and 1 km porous ($\rho = 1.760$ g cm$^{-3}$) and non-porous ($\rho = 2.700$ g cm$^{-3}$) basalt impactors, and 0.5 km and 1 km porous ($\rho = 0.600$ g cm$^{-3}$) and non-porous ($\rho = 0.917$ g cm$^{-3}$) ice impactors. We model 0.5 km bodies in order to sample relatively smaller impactors that satisfy size estimates for both ice and basalt bolides, and we model 1 km diameter impactors for easy comparison to previous SL9 models.  An incident angle of $69^{\circ}$ from the vertical and an impact latitude of 55$^{\circ}$.10 S were used in the simulations \citep{2010ApJ...715L.155S}. The gravitational acceleration at this latitude, including the J2 and centrifugal terms, is 2582 cm s$^{-2}$. An impact velocity of $v$ = 61.4 km s$^{-1}$ was used for the purposes of comparison to previous SL9 simulations \citep{2006ApJ...646..642K,2011ApJ...731....3P}. Several runs were also conducted to test the degree of dynamical chaos, the sensitivity of results to initial conditions, present in simulations with 2009 impact parameters \citep{2006ApJ...646..642K}. 

\begin{deluxetable}{cccccc}
\tabletypesize{\scriptsize}
\tablecaption{Impact Parameters\label{tbl-1}}
\tablewidth{0pt}
\tablehead{
\colhead{Case} & \colhead{Material} & \colhead{Density} & \colhead{Diam.} & \colhead{Angle} & \colhead{Latitude} \\
\colhead{label} & \colhead{} & \colhead{(g/cm$^{3}$)} & \colhead{(km)} & \colhead{} & \colhead{}
 }
\startdata
I05p & Ice & 0.600 & 0.5 & 69$^{\circ}$ & 55$^{\circ}$.10 S \\
I05n & Ice & 0.917 & 0.5 & 69$^{\circ}$ & 55$^{\circ}$.10 S \\
B05p & Basalt & 1.760 & 0.5 & 69$^{\circ}$ & 55$^{\circ}$.10 S \\
B05n & Basalt & 2.700 & 0.5 & 69$^{\circ}$ & 55$^{\circ}$.10 S \\
I10p & Ice & 0.600 & 1.0 & 69$^{\circ}$ & 55$^{\circ}$.10 S \\
I10n & Ice & 0.917 & 1.0 & 69$^{\circ}$ & 55$^{\circ}$.10 S \\
B10p & Basalt & 1.760 & 1.0 & 69$^{\circ}$ & 55$^{\circ}$.10 S \\
B10n & Basalt & 2.700 & 1.0 & 69$^{\circ}$ & 55$^{\circ}$.10 S \\
\tablenotemark{a}SL9p & Ice & 0.600 & 1.0 & $43^{\circ}.09$ & 44$^{\circ}$.02 S \\
SL9n & Ice &  0.917 & 1.0 & $43^{\circ}.09$ & 44$^{\circ}$.02 S \\
\enddata
\tablenotetext{a}{SL9 parameters used in \cite{2011ApJ...731....3P}.}
\end{deluxetable}

\begin{figure*}[t]
\centerline{\includegraphics[width=\textwidth]{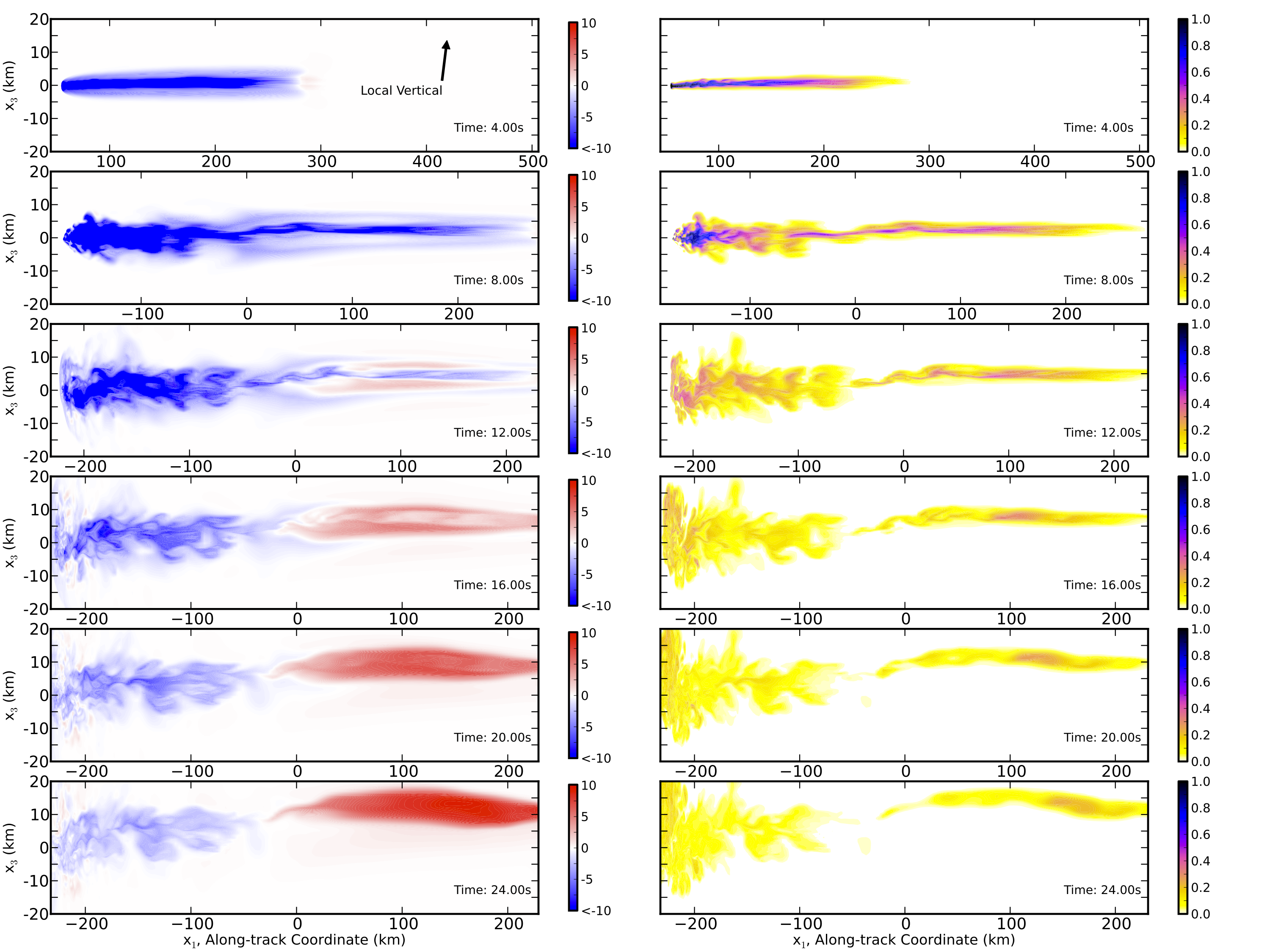}}
\caption{Progression of our impact simulations. Left: Along-track velocity values [km s$^{-1}$], i.e., the velocity in the $x_{1}$ direction. A red color indicates an upward velocity, material moving higher into the Jovian atmosphere; a blue color indicates a downward velocity, material moving deeper into the atmosphere. The red areas are indicators of the rising plumes. Right: the fraction of impactor material compared to Jovian air. The along-track coordinates are given at the bottom of the plots. The I10n case is shown here. The first panel shows an arrow indicating the direction of up in local Jovian coordinates. Note: the $x_{1}$ and $x_{3}$ length scales are not the same.}
\label{12panels}
\end{figure*}

\begin{deluxetable*}{ccccccccccc}
\tabletypesize{\scriptsize}
\tablecaption{Simulation Results\label{tbl-2}}
\tablewidth{0pt}
\tablehead{
\colhead{Case} & \colhead{Terminal Depth\tablenotemark{a}} & \colhead{Pressure at} & \colhead{Total Energy} & \colhead{Maximum Plume Velocity} & \colhead{Pinch-off Location\tablenotemark{a}} & \colhead{Pressure at} \\
\colhead{label} & \colhead{(km)} & \colhead{Terminal Depth} & \colhead{(erg)} & \colhead{(km/s)} & \colhead{(km)} & \colhead{Pinch-off Location } \\
\colhead{} & \colhead{} & \colhead{(bar)} & \colhead{} & \colhead{} & \colhead{} & \colhead{(bar)}
}
\startdata
I05p & -24 & 2.44 & 7.4 $\times$ $10^{26}$ & 7.5 & 21 & 0.365  \\
I05n & -36 & 3.52 & 1.1 $\times$ $10^{27}$ & 9.1 & 19 & 0.408  \\
B05p & -54 & 5.67 & 2.2 $\times$ $10^{27}$ & 8.0 & -1 & 1.06 \\
B05n & -55 & 5.83 & 3.3 $\times$ $10^{27}$ & 8.3 & -2 & 1.10 \\
I10p & -61 & 6.72 & 5.9 $\times$ $10^{27}$ & 10.3 & 4 & 0.891 \\
I10n & -80 & 10.2 & 9.1 $\times$ $10^{27}$ & 12.5 & -4 & 1.19 \\
B10p & -102 & 15.8 & 1.7 $\times$ $10^{28}$  &  11.0 & -18 & 2.00 \\
B10n & -114 & 19.7 & 2.6 $\times$ $10^{28}$  &  11.2 & -23 & 2.37 \\
\tablenotemark{b}SL9p & -124 & 23.5 & 5.9 $\times$ $10^{27}$  & 17.2 & -33 & 3.23\\
SL9n & -150 & 35.0 & 9.1 $\times$ $10^{27}$  & 16.0 & -50 & 5.15\\
\enddata
\tablenotetext{a}{Given in $z$, altitude relative to the 1-bar pressure level.}
\tablenotetext{b}{Results from simulation using parameters from \cite{2011ApJ...731....3P}.}
\end{deluxetable*}

\begin{figure}[h]
\centerline{\includegraphics[width=\columnwidth]{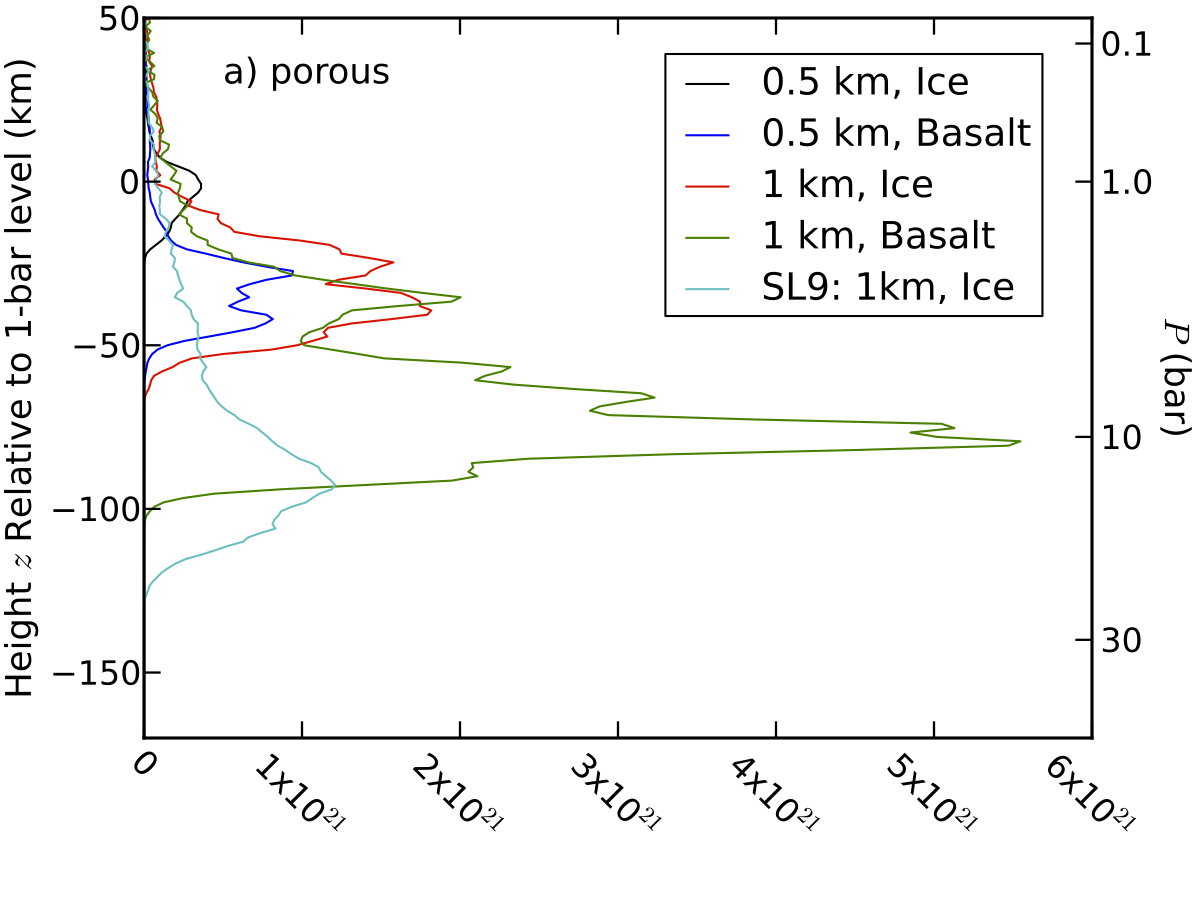}}
\centerline{\includegraphics[width=\columnwidth]{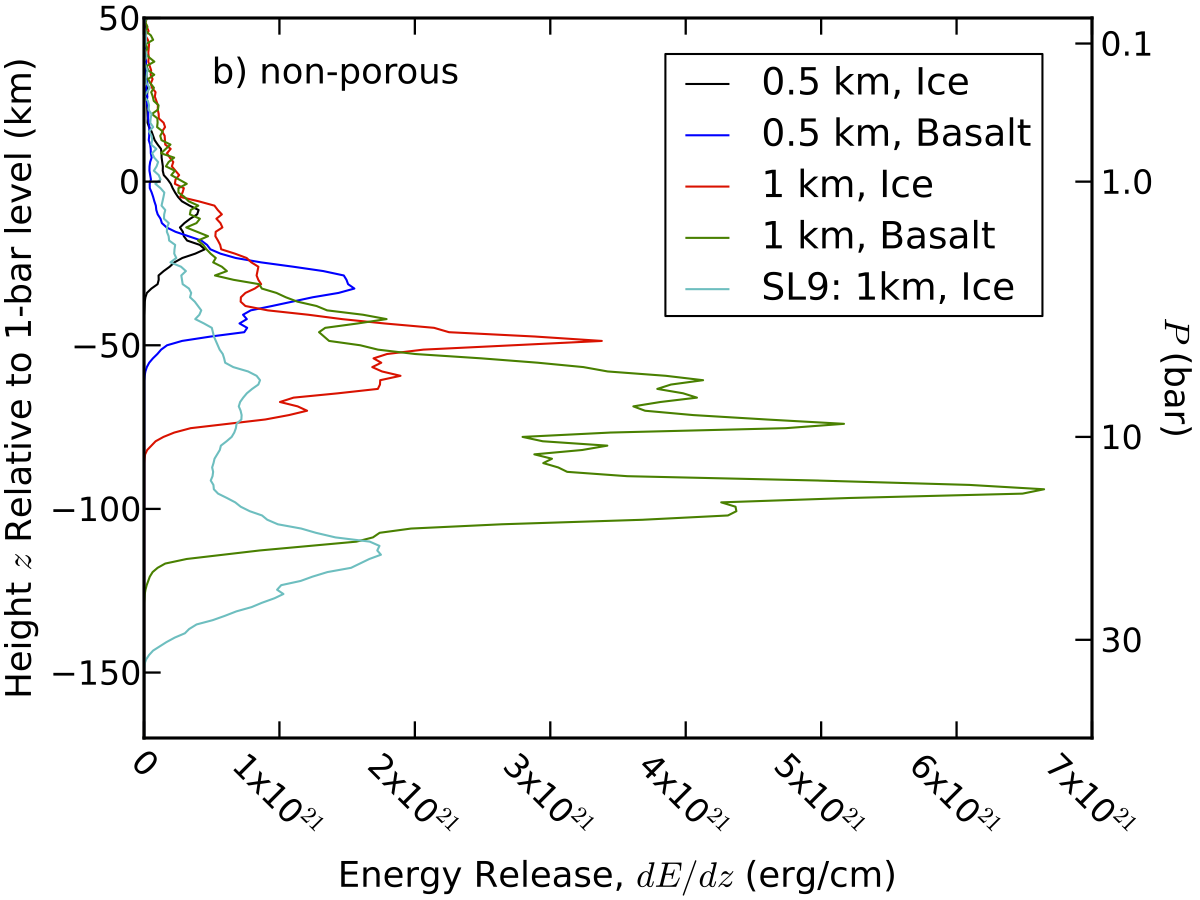}}
\caption{Kinetic energy deposition curves for the porous impact simulations (a) and the non-porous impact simulation (b). This plot demonstrates the increase of both terminal depth and total energy deposited as the size and density of the impactors grow.}
\label{EDcurve_por}
\end{figure}

\begin{figure*}[t]
\centerline{\includegraphics[width=\textwidth]{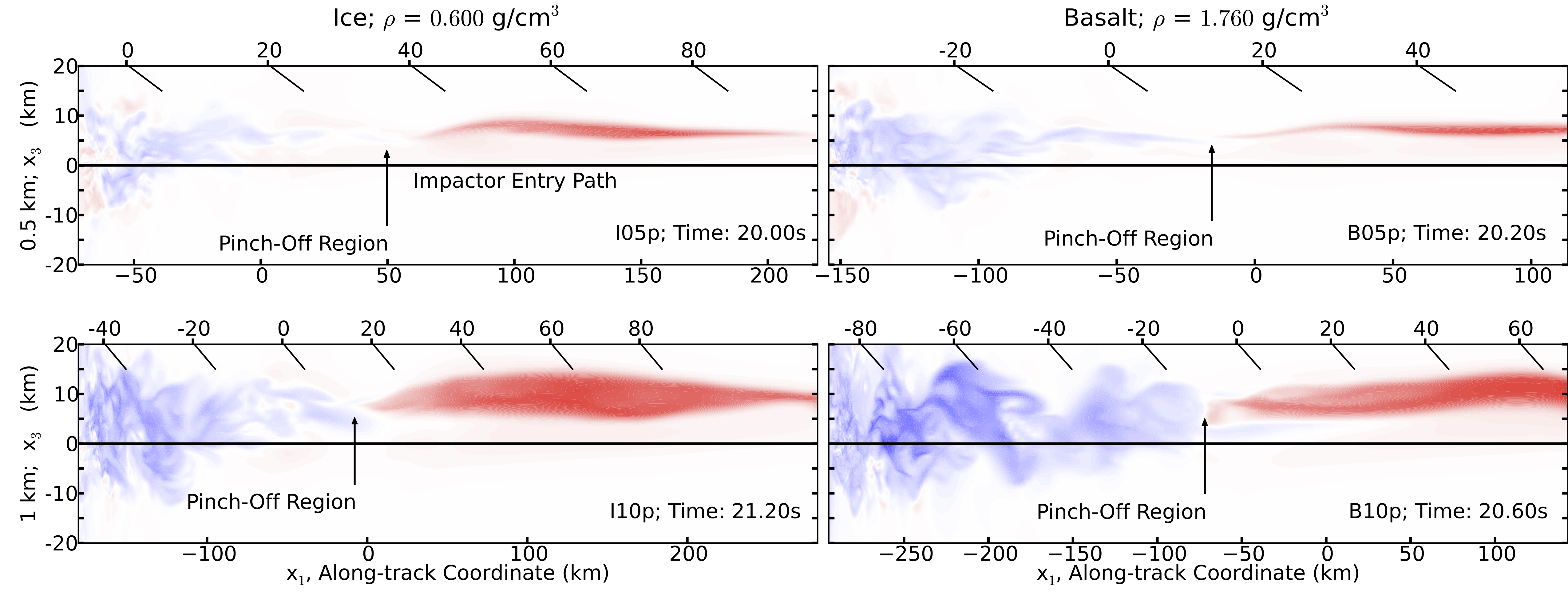}}
\caption{Plume structures of the porous 2009 impact cases. Along-track velocity is shown. The same color bar used for the left panels of Figure \ref{12panels} also applies here. Top-left: 0.5 km, porous ice; Top-right: 0.5 km, porous basalt; Bottom-left: 1 km, porous ice; Bottom-right: 1 km, porous basalt. The along-track coordinates are given at the bottom of the plots. The top of the plots indicates lines of constant height, $z$ [km], in Jupiter's atmosphere. $z$ = 0 km represents the 1 bar pressure level. Note: the $x_{1}$ and $x_{3}$ length scales are not the same.}
\label{comp_por}
\end{figure*}

\begin{figure*}[t]
\centerline{\includegraphics[width=\textwidth]{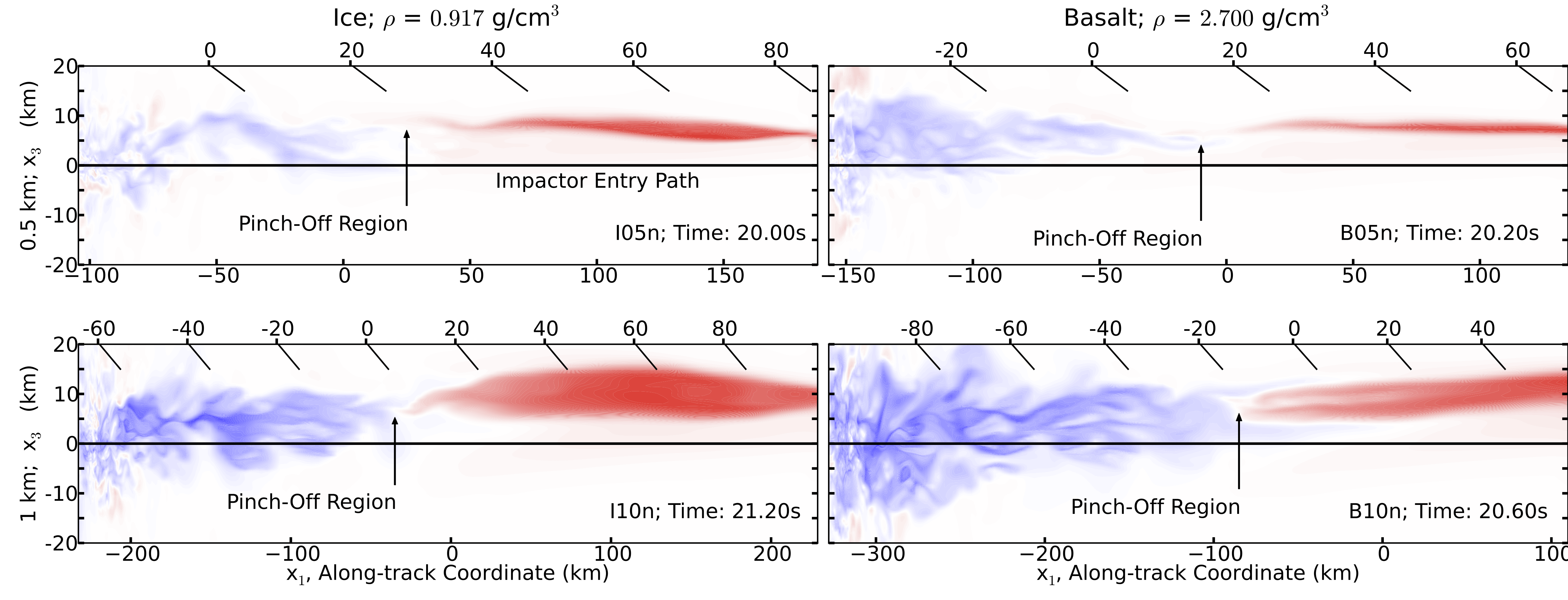}}
\caption{Plume structures of the non-porous 2009 impact cases. Along-track velocity is shown. The same color bar used for the left panels of Figure \ref{12panels} also applies here. Top-left: 0.5 km, non-porous ice; Top-right: 0.5 km, non-porous basalt; Bottom-left: 1 km, non-porous ice; Bottom-right: 1 km, non-porous basalt. The along-track coordinates are given at the bottom of the plots. The top of the plots indicates lines of constant height, $z$ [km], in Jupiter's atmosphere. $z$ = 0 km represents the 1 bar pressure level. Note: the $x_{1}$ and $x_{3}$ length scales are not the same.}
\label{comp_non_por}
\end{figure*}

\begin{figure*}[t]
\centerline{\includegraphics[scale=0.58]{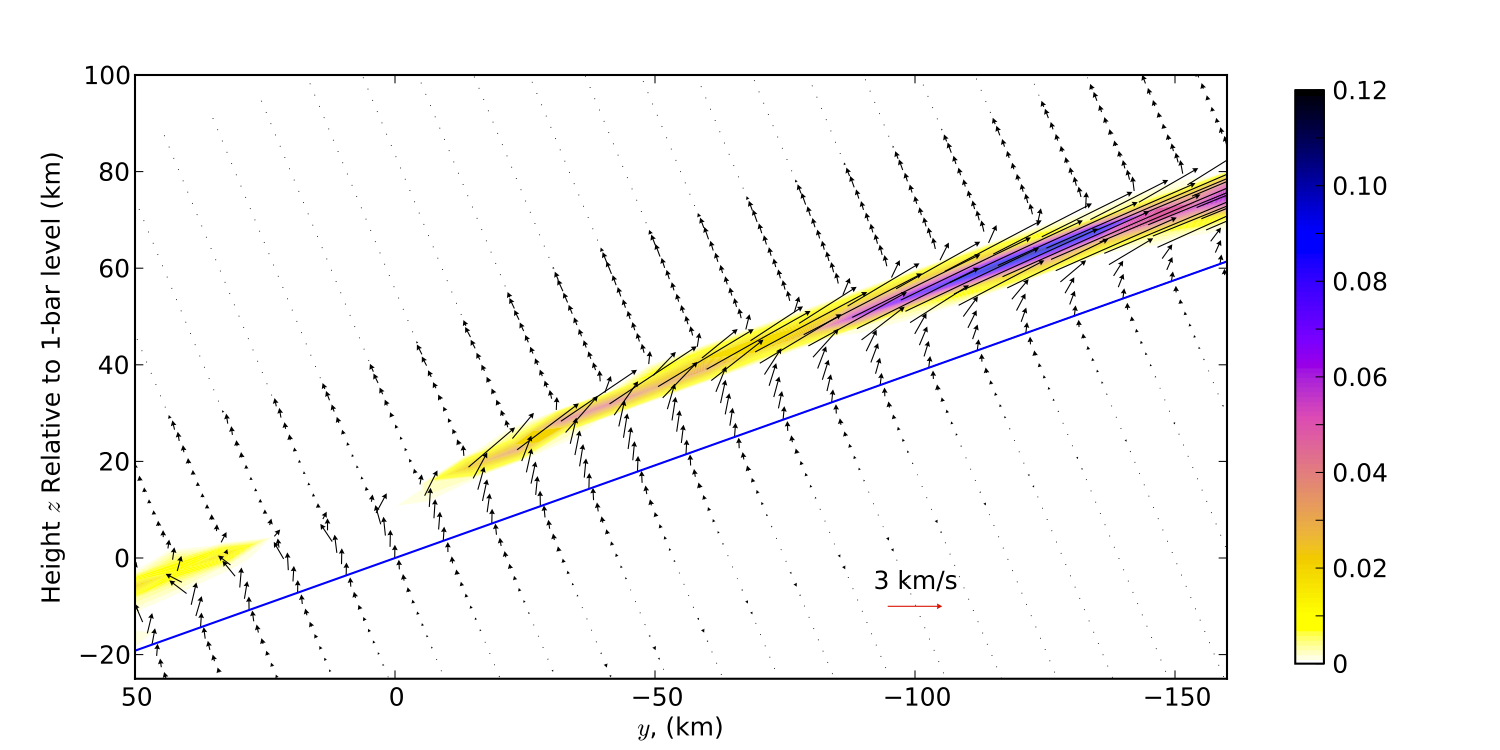}}
\caption{Velocity vector field of the impact plume for case I10p, 27.0 seconds after impact. Color gives the fraction of impactor material present in the rising plume. The straight, dark line indicates the initial trajectory of the impactor.}
\label{2009_vector}
\end{figure*}

\begin{figure*}[t]
\centerline{\includegraphics[scale=0.48]{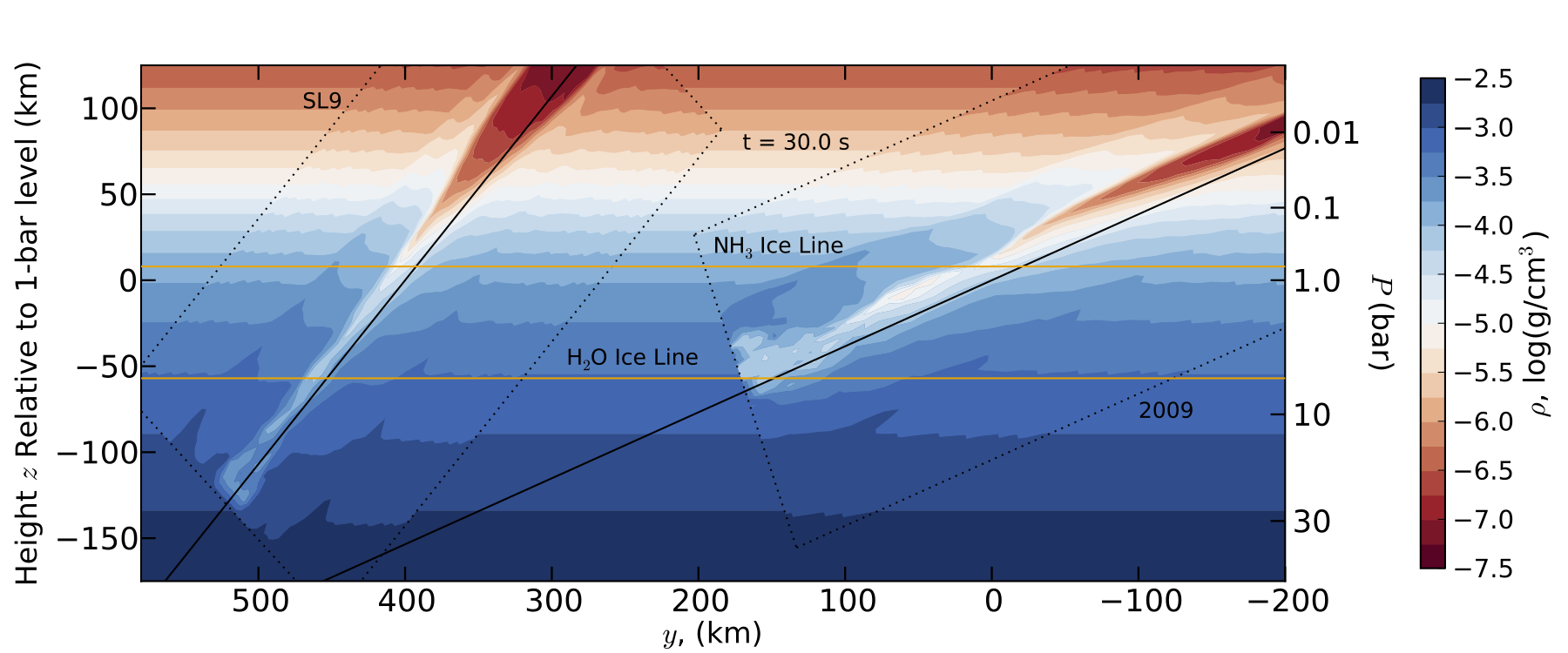}}
\caption{Comparison of the atmospheric density distribution during the aftermath of two impact events, one at the SL9 impact angle (left) and one at the 2009 impact angle (right). The undisturbed Jovian atmospheric density profile is plotted in the background. Contours give the $\log{(\rho)}$ values for each impact and Jupiter's atmosphere. The local vertical is given on the left axis; the corresponding pressure values are given on the right axis. For clarity, the SL9 case is shifted to the left in the $y$ coordinate by 400 km relative to the 2009 case. Both impactors are identical 1 km non-porous ice bodies impacting at 61.4 km s$^{-1}$ (cases SL9p and I10p). The straight, dark lines indicate the initial trajectories of the impactors and the dotted lines indicate the boundaries of the simulations' computational grids.}
\label{comp_i09_SL9}
\end{figure*}

\section{Results}
\label{sec:results}

Impact parameters for all the major cases run in the present paper, plus the parameters used by \cite{2011ApJ...731....3P} for the SL9 case, are given in Table \ref{tbl-1}, and a summary of simulation results is given in Table \ref{tbl-2}. A case label is given to each simulation. Figure \ref{12panels} gives a series of snapshots of the 1 km non-porous ice impactor's simulated decent into the Jovian atmosphere. This figure shows the typical progression of our simulations. The first row of panels, $t = 4.00$ seconds after impact, shows the impactor traveling before it has begun to fall apart. As shown by \cite{2011ApJ...731....3P}, all of the impactor material is contained within a narrow trail following the body, constrained by the shock system on the trailing edge of the impactor. In the second row, $t = 8.00$ seconds, the body has begun significant breakup and decelerates. The impactor material near and around the body begins to spread out quickly, and the shock system becomes turbulent. Material left behind by the bolide in the back half of the grid, within the high-temperature, low-density region that will form the plume, spreads out less rapidly. By $t = 12.00$ seconds, the impactor has become very incoherent and has almost reached its terminal depth. Impactor material and Jovian air in the back region begin their rapid ascent in the atmosphere, signifying the start of plume genesis. The plume is indicated by the growing red region. The majority of the impactor material travels to its terminal depth between $t = 16.00$ and $24.00$ seconds, and the plume can be seen speeding up and rising straight up in the x$_{3}$ direction. As the impactor travels deeper, its vapor diffuses into the Jovian atmosphere, and the impactor fraction decreases with time. The plume increases its blowout speed and rises well above the initial impact trajectory of x$_{3}$ = 0. The impactor material in this region also continues to dissipate as the plume travels in the atmosphere. In general, all the impact simulations presented in this paper proceed as described above and resemble that pictured in Figure \ref{12panels}. Noteworthy differences in the depths reached by the impactors and the development of each plume do exist, however, and are explained in the following subsections.

\subsection{Terminal Depth and Energy Deposition}
Figure \ref{EDcurve_por} shows the kinetic energy deposition curves for the porous and non-porous 2009 impact cases. Also included for comparison on each plot is the energy deposition curve of the appropriate SL9 impact case. These plots, similar to those of \cite{2006ApJ...646..642K}, show the amount of energy an impactor releases to the surrounding atmosphere per unit of altitude traveled. The location of the initial and sudden increase of an energy deposition curve is an indicator of the height at which rapid and extensive structure loss of the impactor begins. Maximum energy deposition occurs when the nucleus of the impactor loses all coherency, and the impactor explodes.

The altitude at which the energy deposition returns to a value of zero is the terminal depth of the impactor. Each of the simulated impactors begins significant breakup at different altitudes and reach varying terminal depths. These terminal depths are listed in Table \ref{tbl-2} and are given in the local vertical coordinate, $z$. A positive $z$ represents an altitude above the 1-bar level in the Jovian atmosphere, and a negative $z$ represents an altitude below the 1-bar level. The terminal depths are consistent with energy considerations: larger, more dense impactors will penetrate the deepest, whereas smaller, less dense impactors will reach shallower depths. The total energies of the impactors, obtained by integrating $dE/dz$, are given in Table \ref{tbl-2}.

\subsection{Plume Development and ``Pinch-off'' Regions}
Figure \ref{comp_por} shows the plume structure at comparable times of plume evolution for each porous 2009 impact simulation. Figure \ref{comp_non_por} shows the same for the non-porous 2009 impact simulations. Within the simulated $\sim$ 30 seconds, higher resultant ejection speeds are reached in the 1 km diameter cases. Markedly smaller plumes and generally lower ejections speeds are seen in the 0.5 km impactor cases.

Figure \ref{2009_vector} shows the velocity distribution of a rising plume and gives the fraction of impactor material contained within. All of the plume ejecta are above the initial impact trajectory. This is different from the SL9 simulations, where the plume expands to a greater diameter in the simulated time and still crosses the initial impact path \citep{2011ApJ...731....3P}. The distribution of impactor material within the plume is similar to the SL9 simulation conducted by \cite{2011ApJ...731....3P}, however. A maximum of $\sim$ 10\% impactor material is located near the top of the ejecting plume, and this fraction decreases as one moves deeper down the length of the plume. The  ejection angles of the plumes are similar across all the 2009 cases, and within the simulated $\sim$ 30 seconds, the plumes attain  ejection angles of 60$^{\circ}$-70$^{\circ}$ from the vertical.

Just as in the SL9 simulations conducted by \cite{2011ApJ...731....3P}, a pinch-off region appears in each of the 2009 cases. These regions are shown in Figures \ref{comp_por} and \ref{comp_non_por}. Above this pinch-off level, heated Jovian atmosphere containing a small fraction of impactor material rises and expands as a plume (shades of red in Figures \ref{comp_por} and \ref{comp_non_por}); below this level, a majority of the impactor material continues downward (shades of blue in Figures \ref{comp_por} and \ref{comp_non_por}) and will later rise more slowly and independently of the ejecting plume as a bubble-like region of impactor material. The approximate location of this pinch-off region for each case is listed in Table \ref{tbl-2}. 

The altitudes at which the pinch-off regions occur appear to correspond with the altitudes at which the impactor begins rapid and extensive structure loss \citep{2011ApJ...731....3P}. We are currently working to better characterize and garner additional details about the relationship between the pinch-off region and impactor break-up.

\section{Discussion}
\label{sec:disc}

\subsection{Observations and Simulations}
Observations conducted by \cite{2010Icar..210..722D}, \cite{2010A&A...524A..46F},  \cite{2010ApJ...715L.150H}, \cite{2011Icar..211..587O}, and \cite{2010ApJ...715L.155S} place several constraints on the impact parameters and the dynamics of the 2009 impact. In this section, we will examine these constraints and compare them to the results of the present simulations.

Two interesting constraints include the extent of debris distribution and temperature perturbations in the Jovian atmosphere as a result of the impact. Debris deposition was constrained roughly between the 10-mbar and 300-mbar levels  \citep{2010Icar..210..722D, 2010ApJ...715L.150H, 2011Icar..211..587O}. Thermal perturbations reached higher up in the atmosphere than the debris, up to about the 0.1--20-mbar levels, and were present slightly deeper into the atmosphere than the debris, down to the 400-600 mbar levels \citep{2010Icar..210..722D, 2011Icar..211..587O}. A lack of excess methane emission in the upper stratosphere is also an indicator of heating limited to pressures greater than 10 mbar \citep{2010A&A...524A..46F}. This is very different from the the SL9 impacts, in which plumes reached thousands of kilometers above the 100-mbar level in Jupiter's atmosphere \citep{HammelEtal1995SL9HST,2000Icar..146...19J}. 

Enhanced levels of ammonia in the lower stratosphere also place boundaries on the penetration depth of the plume.  \cite{2011Icar..211..587O} found that a vertical profile of ammonia peaking in the 20--30-mbar region is required to reproduce the spectral shape of the 2009 NH$_{3}$ emission at the impact location. This enhanced presence of ammonia in the Jovian stratosphere implies the 2009 impact wake probably reached down to the 600-700 mbar levels, retrieving tropospheric ammonia and transporting it to the stratosphere as the plume rose \citep{2010Icar..210..722D, 2011Icar..211..587O}.  However, it is likely that the downward jet that would form the plume did not penetrate much farther than the 700-mbar level since the NH$_{3}$ gas was contained close to the center component of the impact streak \citep{2010Icar..210..722D}, and the jet probably did not make it to the Jovian water cloud level since shock chemisty of the 2009 event favored production of ethane and other hydrocarbons over CO and H$_{2}$O \citep{2010A&A...524A..46F}.

Table \ref{tbl-2} gives the pressure levels at which the pinch-off regions occur for each of the conducted simulations. The pinch-off region is a measure of the deepest penetration of the fireball that forms the plume and is the lower limit from which material may be dredged up from the Jovian depths. In general, the plume jet is small and weak at the pinch-off region,  but grows in speed as one moves up the plume channel. Cases I05p and I05n have pinch-off regions located at $\sim$365-mbar and $\sim$410-mbar, respectively. Ice impactors of these sizes, densities, and impact velocity result in plumes that do not penetrate down to the stratospheric ammonia reservoir. Cases B05p and B05n result in relatively small plumes whose pinch-off regions penetrate down to the $\sim$1.10-bar level, past the top levels of Jupiter's NH$_{3}$ clouds, and speeds in the lower plume jet around the 700-mbar level reach  $\sim$1.8 km s$^{-1}$ for B05p and  $\sim$1.0 km s$^{-1}$ for B05n.  Cases I10p and I10n have pinch-off regions located at $\sim$890-mbar and $\sim$1.2-bar, respectively. For these two 1 km ice cases, speeds in the lower plume jet at $\sim$700-mbar reach about 1.8 km s$^{-1}$ for I10p and 2.8 km s$^{-1}$ for I10n. Cases B10p and B10n result in larger plumes that penetrate well past 700-mbar, down to pressures around the 2 and 3 bar levels, respectively, but stay above the H$_{2}$O cloud tops around 6 bar. At the $\sim$700-mbar level, speeds in the lower plume jet reach about 4.6 km s$^{-1}$ for B10p and about 5.6 km s$^{-1}$ for B10n.

Cases I05p and I05n do not satisfy the observational constraints, so it is likely that bodies with these impact characteristics did not cause the atmospheric response seen on Jupiter in July 2009, setting a lower limit on the possible size, density, and impact velocity of the bolide. The plumes of cases B05p, B05n, I10p, I10n, B10p, and B10n penetrate deep enough to reach the ammonia clouds in the Jovian atmosphere, and so these impactors remain possible 2009 candidates. Further constraining the possibilities of the 2009 impact will require continued exploration of the possible impact parameter space, detailed ammonia transport calculations, and extending simulations both spatially and temporally.

\cite{2011Icar..211..587O} give $7 \pm 2 \times 10^{26}$ erg as a lower estimate for the energy of the impact. This is a lower estimate because it does not take into account the 4 days of cooling that passed between the impact and the observations nor the large amounts of energy lost to other dynamical processes in the impact, such as plume formation and the transport of atmospheric waves. All of the present simulations satisfy this lower limit.

\begin{figure}[t]
\centerline{\includegraphics[width=\columnwidth]{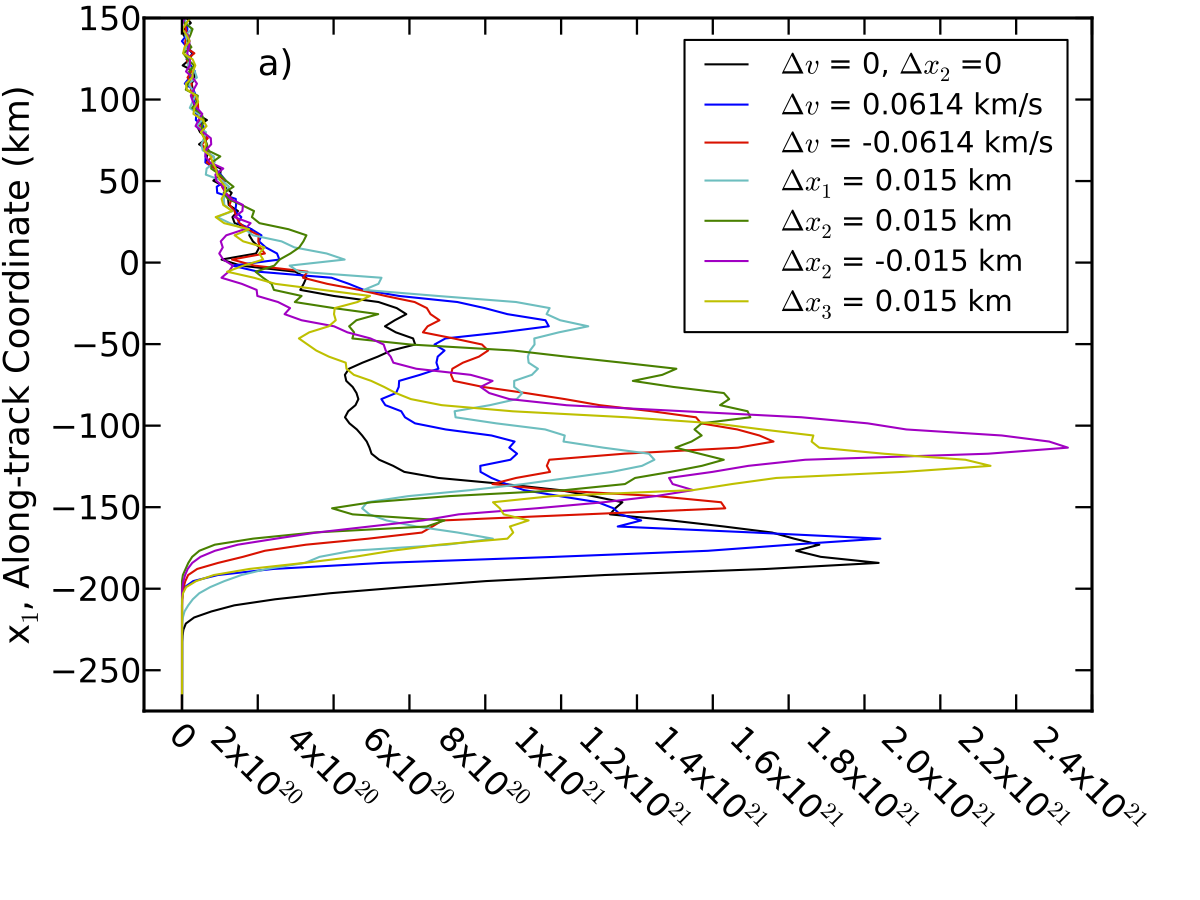}}
\centerline{\includegraphics[width=\columnwidth]{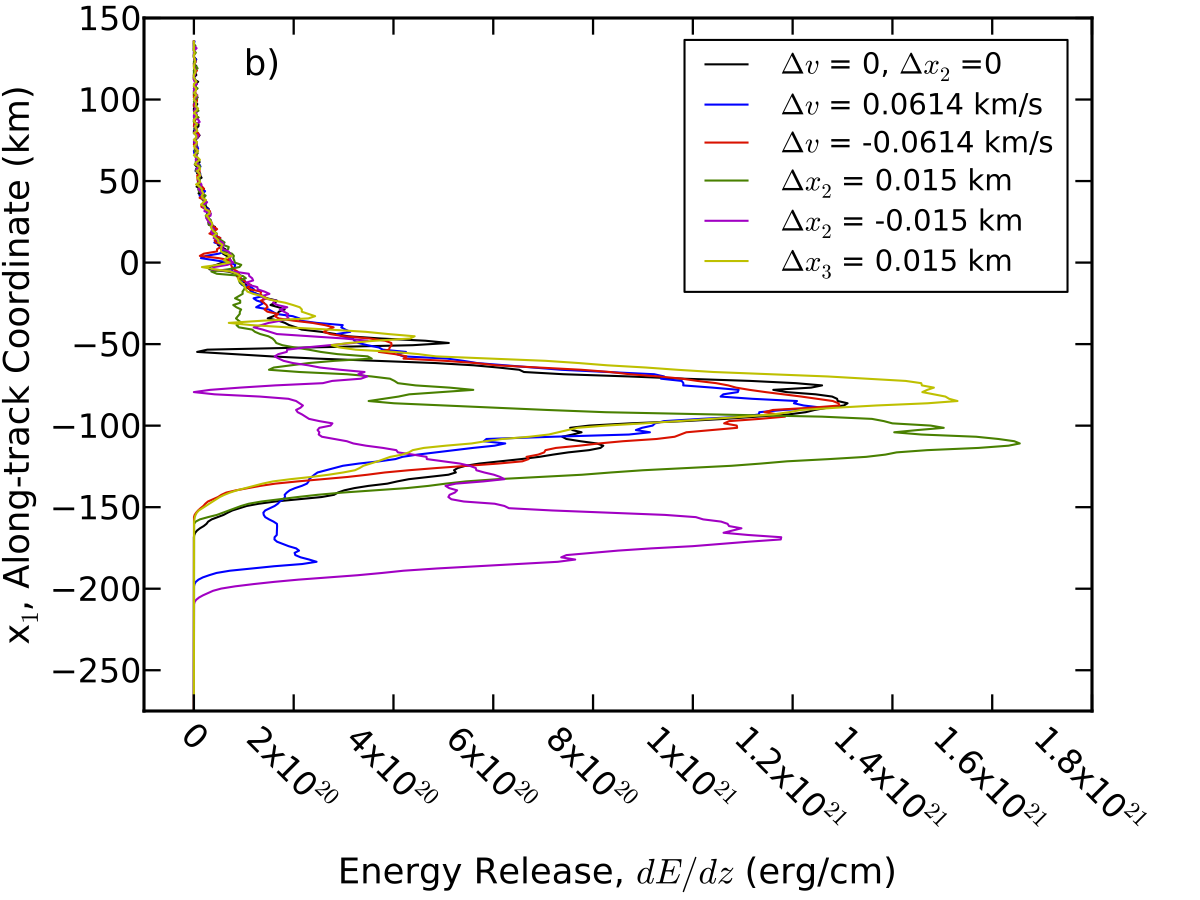}}
\caption{Dynamical chaos analysis. Kinetic energy deposition curves at the 2009 incident angle (a) and the SL9 incident angle (b). The black line gives the energy deposition for the nominal case of  a 1 km porous ice impactor traveling at $v = 61.4$ km s$^{-1}$. The bottom plot is Figure 7 (d) from \cite{2006ApJ...646..642K}.}
\label{chaoscurve}
\end{figure}

\subsection{2009 vs. SL9 Impacts}
Figure \ref{comp_i09_SL9} shows a snapshot of the aftermath of both an SL9 impact and a 2009 impact. The figure gives the density distribution for both cases, plotted against the undisturbed Jovian density distribution. Case SL9p and case I10p are used in the figure, i.e., the impacts in this figure are identical, except for the incident angle and the latitude at which the impactor strikes Jupiter. Both cases exhibit features characteristic of all our present impact simulations. At the terminal depths of the impactors, a relatively-low-density region containing most of the impactor material exists. This is the bubble-like region below the pinch-off level seen in the simulations of \cite{2011ApJ...731....3P}. Shocks can be seen propagating away from the impact path, and at the boundaries of plume formation, two shock waves can be seen. The lowest-density region marks the location of the  plume. In the upper atmosphere, the plume gains speed, rises, and expands; in the lower atmosphere, the plume becomes slender, tapering to the pinch-off region. Though both impactors are identical and share general characteristics, the terminal depths and plume development are strikingly different at the differing incident angles.

In this particular comparison, the terminal depth of the SL9 impactor is $\sim$ 60 km deeper than the 2009 impactor's terminal depth. This is mainly due to the larger incident angle of the 2009 impact: for a given distance traveled along the path of each impactor (in $x_{1}$), displacement in height $z$ will be smaller at higher incident angles. The pinch-off regions differ by a little more than a scale height, with the SL9 pinch-off reaching to $\sim$ 30 km below the 1-bar level and the 2009 pinch-off occurring $\sim$ 4 km above the 1-bar level. There is also a distinct difference between the developing plumes. For the SL9 case, the plume begins to expand significantly as it rises in the atmosphere. The SL9 plume reaches a diameter of $\sim$35 km by 30 seconds after impact. Such rapid expansion is not seen in the 2009 plume, however. During the formation of the 2009 plume, it rises above the impact path while undergoing relatively little expansion. By 30 seconds after impact, the 2009 plume only reaches a diameter of $\sim$15 km. The difference in plume sizes can be seen in Figure \ref{comp_i09_SL9} which shows the SL9 plume about twice the size of the 2009 plume. Table \ref{tbl-2} reveals that the SL9 plume contains air and entrained impactor material traveling at speeds about 7 km s$^{-1}$ greater compared to the 2009 plume. The larger plume and heightened speeds in the SL9 case relative to that of 2009 implies a dependance of plume size and maximum plume height on incident angle.

\subsection{Dynamical Chaos}
In order to observe the sensitivity of the present results to initial conditions, we conducted several 1 km porous ice impactor simulations with slightly different initial conditions compared to a nominal case. These differences in initial conditions include a change in the impact velocity ($\Delta v$) by 0.1\% and a shift in the initial position of the impactors (in each direction $x_{1}$, $x_{2}$, and $x_{3}$) by a half a grid cell, $\sim$ 15 m in this case. The variation in terminal depths and peak energy deposition locations is a measure of the dynamical chaos present in the simulations. \cite{2006ApJ...646..642K} conducted this same analysis and found that among the four SL9 impact cases they simulated, a 1 km porous ice impactor was subject to the highest degree of dynamical chaos. Only slightly changing initial conditions for this case resulted in relatively large variations in the terminal depths and peak energy deposition locations.

Figure \ref{chaoscurve} (a) gives our chaos analysis of the 2009 impact for the porous ice case. Comparing Figure \ref{chaoscurve} to Korycansky et al.'s (2006) Figure 7 (d), seen here as our Figure \ref{chaoscurve} (b), we do not find the dynamical chaos to be significantly different at an incident angle of $\theta = 69^{\circ}$. The standard deviation of the terminal depths at the 2009 impact angle is $S_{i09} \approx$ 13.2 km in the along-track coordinate, and the range in terminal depths is about 44 km. At the $\theta \approx 45^{\circ}$ SL9 angle of incidence, the standard deviation of the terminal depths is  $S_{SL9} \approx$ 20.8 km in the along-track coordinate, and the range in terminal depths is about 53 km. 

We conducted a simple F-Test to compare the variation in terminal depths between the 2009 and SL9 cases. Our test $F$ statistic, $F = S^{2}_{SL9} /S^{2}_{i09} = 2.48 $, was compared to the critical $F$ statistic at the 5\% level, $F = 4.39$. Though visually the terminal depths for the 2009 porous ice case seem less scattered, and the range and standard deviation of these terminal depths is less at the 2009 incident angle, we cannot say that the standard deviation of the terminal depths at the SL9 incident angle is significantly larger than at the 2009 incident angle because our test $F$ statistic was not greater than the critical value.

\section{Conclusions}
\label{sec:con}
We have presented several possibilities for the impact and immediate aftermath of the 2009 collision into the Jovian atmosphere. The results of the present simulations provide insight into the impact event and also provide information about the variation in atmospheric response due to changes in impact parameters. At the estimated 2009 incident angle of $\theta = 69^{\circ}$, we see several differences between plumes generated from 0.5 km impactors and 1 km impactors. Within the simulated $\sim$ 30 seconds, the 0.5 km impactor events produce relatively smaller and slower plumes while the 1 km impactor events produce relatively larger and faster plumes. 

The penetration depths of the impactors and the pinch-off regions are associated with the nature of the impactors: at a given incident angle, the larger the impactor and the heavier the impactor material, the deeper the locations of the terminal depth and the pinch-off region. Dynamical chaos present at the 2009 incident angle for the 1 km porous ice impactor did not prove to be significantly less than that of the most chaotic impactor case given by \cite{2006ApJ...646..642K}.

Comparing the aftermaths of an SL9 impact and the 2009 event reveals several differences that may have consequences for the observable manifestation of Jupiter's atmospheric response. The impact plume produced at the SL9 incident angle is significantly larger and faster than that of the impact plume produced at the 2009 incident angle.

Given observations of thermal perturbations, debris deposition, and ammonia transport in the Jovian atmosphere after the 2009 event, constraints have been placed on the possible outcomes of the impact, including plume speeds and pinch-off region locations \citep{2010Icar..210..722D, 2010A&A...524A..46F, 2010ApJ...715L.150H, 2011Icar..211..587O}. Of the eight cases considered in the present paper, the 0.5 km ice cases cannot explain the atmospheric disturbance observed after the 2009 impact; these impactors' plume jets do not penetrate deep enough to explain stratospheric NH$_{3}$ observations \citep{2010Icar..210..722D, 2011Icar..211..587O}. We thus set a lower limit on the size and density of the 2009 impactor. All 1 km impactor plumes and the 0.5 km basalt impactor plumes reach down to the ammonia ice cloud level in Jupiter's troposphere. To better constrain the impact characteristics of the 2009 event, more simulations are required spanning more of the possible parameters proposed by observations. Ammonia transport from the upper troposphere to the stratosphere by means of the rising plume jets must also be looked at in more detail. Through this analysis, we will narrow the range of possible impactor circumstances that produced the atmospheric disturbance on Jupiter in 2009. 

\indent

This work was supported by National Science Foundation Grants AST-0813194, AST-0964078, and AST-1109729 and NASA Planetary Atmospheres Program Grant NNX11AD87G.

\bibliography{pond-et-al-2011-2009-impact}

\end{document}